\documentclass[letterpaper,prl,twocolumn,superscriptaddress,preprintnumbers,showpacs]{revtex4}

\usepackage{graphicx}
\usepackage{bm}
\usepackage{amssymb, amsmath}

\newcommand{\mysw}[1]{\scriptscriptstyle #1}

\def\be{\begin{eqnarray}}
\def\ee{\end{eqnarray}}

\newcommand{\Eq}[1]{Eq.~(\ref{#1})}

\newcommand{\p}{\partial}

\newcommand{\f}[2]{\ensuremath{f^{#1}_{#2}}}
\newcommand{\coup}[2]{\ensuremath{c^{#1}_{#2}}}
\newcommand{\Coup}[2]{\ensuremath{\zeta^{#1}_{#2}}}
\newcommand{\tf}[2]{\ensuremath{{\tilde f}^{#1}_{#2}}}
\newcommand{\tc}[2]{\ensuremath{{\tilde c}^{#1}_{#2}}}
\newcommand{\Fig}[1]{Fig. \ref{#1}}

\begin{document}

\title{Phonons in Hubbard ladders studied within the framework of the one-loop renormalization group}
\author{Alexander Seidel}
\affiliation{Department of Physics, University of California, Berkeley, CA 94720}
\affiliation{Materials Sciences Division, Lawrence Berkeley National Laboratory, Berkeley, CA 94720}
\author{Hsiu-Hau Lin}
\affiliation{Department of Physics, National Tsing-Hua University, Hsinchu 300, Taiwan}
\affiliation{Physics Division, National Center for Theoretical Sciences, Hsinchu 300, Taiwan}
\author{Dung-Hai Lee}
\affiliation{Department of Physics, University of California, Berkeley, CA 94720}
\affiliation{Materials Sciences Division, Lawrence Berkeley National Laboratory, Berkeley, CA 94720}

\begin{abstract}
We study the effects of phonons in $N$-leg Hubbard ladders within the framework of a one-loop renormalization group. In particular, we explicitly demonstrate that the role of phonons changes {\em qualitatively} even in the simplest two-leg ladder, as compared to the single-chain system where phonons always dominate. Our numerical results suggest that in the spin-gapped phase of the two-leg ladder, the opening of the spin gap by electron-electron interaction also drives
the electron-phonon interaction to strong coupling, but in a {\em subdominant} fashion. Therefore, even though the inclusion of phonons does not alter the phase, their subdominant relevance strongly renormalizes some physical properties below the energy scale of the spin gap. This might shine some light on the recent experiments showing an anomalous isotope effects in high-temperature superconductors.
\end{abstract}
\maketitle

How relevant phonons are to the high-temperature superconductors
has been an issue of considerable controversy. At one extreme, it
is believed that phonons are irrelevant to the physics of the
cuprates which is dictated by electronic correlations. At the other, phonons are thought to be the glue for the Cooper pairing. Recent experiments of angle-resolved photoemission experiments indicate the significance of electron-phonon coupling in the cuprates\cite{lanzara,shen,gweon} and spark the interests in the controversial role played by phonons again. This is particularly so in the isotope-ARPES experiment by Gweon et al.\cite{gweon} There, the effect of isotope substitution is significantly enhanced in
the superconducting state where the spin gap has opened.\\
\indent
The purpose of this work is to shed some light on the 
effect of electron-phonon couplings in correlated 
electron systems. The challenge here is to do a 
calculation which is {\em controlled} when appropriate
limits are taken, and at the same time is {\em unbiased}
in the sense that no particular scenario is favored
or implicitly assumed, e.g., by a certain choice of
mean field. For the particular case of the 
high-$T_c$ materials, no calculational method is currently 
available that would satisfy both criteria. In the broader
context of correlated electron systems, however, we feel
that the class of ladder systems especially lends itself
to an investigation of the issues stated above: 
Only in ladder systems it is known that purely repulsive electronic
interactions can lead to a spin-gapped phase (with approximate $d$-wave symmetry), in resemblance to scenarios that have been proposed for the superconducting cuprates. If anything, this choice of
system will take the bias off the phonons as the driving mechanism 
behind such a phase, since it already exists without their help.
Moreover, this nontrivial, strongly correlated phase with
various competing order parameters\cite{Rice03}
can be accessed at arbitrarily weak strength of the electronic couplings, 
making it possible to study this problem by means of the perturbative renormalization group (RG). \\
\indent
While the equations we derive are valid for general $N$-leg Hubbard
ladders, we focus on the simplest two-leg ladder in this
paper. Our results demonstrate a clear distinction of the
interplay between electronic and lattice degrees of freedom, depending on whether a finite spin gap develops in the ground states. In the spin-gapped phase, the electron-phonon couplings are always relevant. In particular,
they are driven to large values at the energy scale
of the spin gap, which is set by the electronic interactions.
This happens even
in the limit of infinitesimal bare values of 
electron-phonon couplings. In gapless phases, on the other
hand, electron-phonon couplings are predicted to be irrelevant
at least for sufficiently small bare strength. This dichotomy
indeed bears some resemblance to the experiments on the 
cuprates.\cite{gweon} At the same time, it is found that
the electron-phonon couplings, although relevant, remain
{\em subdominant} compared to interelectronic couplings.
In simple terms, while the phonons do
not {\em drive} the spin-gapped phase, they do {\em dress} it up.
We note that this result is well distinct from that
for the spin-gapped phase in the single chain,
where even
very weak attractive retarded couplings (resulting from
electron-phonon interactions) will eventually dominate over the
nonretarded (electronic) ones and play the essential role of generating the
spin gap.\cite{KIVELSON} 
Our findings in the Hubbard ladder thus seem to set an interesting
balance between the two extremes mentioned previously. On the one
hand, phonons are irrelevant in regard to the pairing mechanism. On
the other, they do strongly influence what we observe in
experiments, e.g., the isotope effects. While the above picture is
suggestive for future explorations in the cuprates, we do warn the
readers that our analysis is confined to ladder systems, which cannot 
be straightforwardly generalized to correlated systems in
higher dimensions.\\
\indent
We now  turn to the rigorous renormalization group analysis of ladder systems. In this Communication, we consider the effect of phonons in the $N$-chain Hubbard ladder by including retarded interactions into the RG treatment. In the following we shall assume general {\it incommensurate} band filling and {\it weak}
interactions. Under these conditions, the effective field theory
for the $N$-chain ladder is a $2N$-species massless Dirac theory
with four-fermion interactions.
The Hamiltonian density ${\cal H} = {\cal H}_{0}+ {\cal H}_{I}$ of the effective theory is given by \cite{FABRIZIO1,BALFISHER, LINFISHER, SCHULZ}:
\begin{eqnarray}
{\cal H}_0=\sum_{i\alpha}
\psi^{\dag}_{\mysw{R}i\alpha}(-iv_{i}\partial_x)\psi^{}_{\mysw{R}i\alpha}
+\psi^{\dag}_{\mysw{L}i\alpha}(iv_{i}\partial_x)\psi^{}_{\mysw{L}i\alpha},
\hspace{1cm}
\nonumber\\
{\cal H}_{I}= \sum_{ij\alpha\beta}
\bar f^{\ell}_{ij} \psi^\dag_{\mysw{R}i\alpha} \psi^\dag_{\mysw{L}j\beta}
\psi^{}_{\mysw{R}i\beta} \psi_{\mysw{L}j\alpha}
+\bar f^{s}_{ij} \psi^\dag_{\mysw{R}i\alpha}\psi^\dagger_{\mysw{L}j\beta}
\psi^{}_{\mysw{L}j\beta} \psi^{}_{\mysw{R}i\alpha}
\nonumber\\
+\bar c^{\ell}_{ij} \psi^\dag_{\mysw{R}i\alpha} \psi^\dag_{\mysw{L}i\beta}
\psi^{}_{\mysw{R}j\beta} \psi^{}_{\mysw{L}j\alpha}
+\bar c^{s}_{ij} \psi^\dag_{\mysw{R}i\alpha} \psi^\dag_{\mysw{L}i\beta}
\psi^{}_{\mysw{L}j\beta} \psi^{}_{\mysw{R}j\alpha} \hspace{8mm}
\label{H}
\end{eqnarray}
where open boundary conditions transverse to the chain direction
are assumed. 
Here, $R$ and $L$ stand for right and left moving electrons,respectively,
$i$ and $j$ are the band indices, and $\alpha,\beta$ are the spin
indices. The superscripts $\ell$ and $s$ of the interaction
parameters refer to large and small momentum transfer scatterings.
Following Ref.\cite{LINFISHER}, the letters $f$ (``forward'') and
$c$ (``Cooper'') are used to distinguish whether the band indices
are conserved at the scattering vertices. For $i=j$, the two
become identical, and we choose $c^{\ell,s}_{ii} \equiv 0$ to avoid
double counting (different from the choice in Ref.\onlinecite{LINFISHER}).
All nonzero interaction parameters are symmetric
in band indices. In addition, terms involving {\it only} the $R$
or the $L$ fermion operators do not participate in the RG
at one-loop order, hence they are omitted.
For the Hubbard ladder, the {\it bare} values of {\it all}
interaction parameters are equal to the on-site repulsion $U$.
 The one-loop RG recursion relations for the interaction parameters
in \Eq{H} are
known,\cite{LINFISHER}
\begin{widetext}
\begin{equation}\label{nrflow}
\begin{split}
\p_l \f{\ell}{ij} &= 2[-(\f{\ell}{ij})^2-(\coup{\ell}{ij})^2+\coup{\ell}{ij}\coup{s}{ij}] -\delta_{ij}\sum_{k\neq i}2\alpha_{ii,k}\,\coup{\ell}{ik}\coup{s}{ik},
\hspace{5mm}
\p_l\f{s}{ij} = (\coup{s}{ij})^2-(\f{\ell}{ij})^2
-\delta_{ij}\sum_{k\neq i} \alpha_{ii,k}\left[(\coup{\ell}{ik})^2+(\coup{s}{ik})^2\right],
\\
\p_l\coup{\ell}{ij} &= -\sum_k\alpha_{ij,k}(\Coup{\ell}{ik}\Coup{s}{kj}+\Coup{\ell}{jk}\Coup{s}{ki})
-4\f{\ell}{ij}\coup{\ell}{ij} + 2\f{\ell}{ij}\coup{s}{ij} + 2\f{s}{ij}\coup{\ell}{ij},
\hspace{5mm}
\p_l\coup{s}{ij} = -\sum_k \alpha_{ij,k}(\Coup{\ell}{ik}\Coup{\ell}{kj}+\Coup{s}{jk}\Coup{s}{ki})
+2\f{s}{ij}\coup{s}{ij},
\end{split}
\end{equation}
\end{widetext}
where $\Coup{\ell,s}{ij}=\coup{\ell,s}{ij}+\delta_{ij}\f{\ell,s}{ii}$ for notational convenience. In the above $\p_l\equiv d/dl$, with $l$ being the logarithm of the ratio between the bare momentum cutoff $\Lambda_0$ and the running cutoff $\Lambda$. Dimensionless couplings
$g_{ij} = \bar{g}_{ij}/\pi(v_i+v_j)$ have been introduced with $g = f^{\ell}, f^{s}, c^{\ell}, c^{s}$ and $\alpha_{ij,k}\equiv(v_i+v_k)(v_j+v_k)/[2v_k(v_i+v_j)]$.

The numerical integration of \Eq{nrflow} shows that quite
generally for the $N$-chain system,\cite{LINFISHER} there exists a
scale $l_d$ where the RG flows diverge due to the opening of spin
or charge gaps in some/all partially filled bands. To ensure the
applicability of the perturbative results, the flows are usually
terminated at a scale $l_c$ (close to but less than $l_d$), where the largest
couplings become of order $1$. Physically, the scale $l_c$ is
identified with the energy scale at which charge/spin gaps open.

To account for the effects of phonons we include retarded four-fermion interactions in \Eq{H}. The latter are effective interactions which are formally identical to those displayed in \Eq{H}, except that a cutoff
$\omega_D$ is imposed for frequency transfer at the retarded
vertices.
By standard momentum shell RG,\cite{SHANKAR} one can derive RG equations for both nonretarded and retarded couplings before the momentum cutoff $\Lambda$ reaches $\omega_{\mysw{D}}/v$ at a scale $l={l}_{\mysw{D}}$, using the method of Refs. \onlinecite{ABRAHAMS, CARON, ZIMANYI, BINDLOSS}.
Below this scale, the retarded couplings are then regarded as
additive renormalizations to the nonretarded ones. 

For $l<{l}_{\mysw{D}}$, the following flow equations are obtained
for the retarded interactions:
\begin{equation}\label{rflow}
\begin{split}
\p_l{\tf{\ell}{ij}}=&-4\f{\ell}{ij}{\tf{\ell}{ij}}+2\f{s}{ij}{\tf{\ell}{ij}}-2({\tf{\ell}{ij}})^2\\
& -4\coup{\ell}{ij}{\tc{\ell}{ij}}+2\coup{s}{ij}{\tc{\ell}{ij}}-2({\tc{\ell}{ij}})^2\\
\p_l{\tc{\ell}{ij}}=&-4{\tf{\ell}{ij}}\coup{\ell}{ij}-4\f{\ell}{ij}{\tc{\ell}{ij}}-4{\tf{\ell}{ij}}{\tc{\ell}{ij}}\\
&+2{\tf{\ell}{ij}}\coup{s}{ij}+2\f{s}{ij}{\tc{\ell}{ij}},
\end{split}
\end{equation}
whereas the nonretarded flow equations \eqref{nrflow} remain unaltered. In the above, we use the notation $\tilde g$ for the retarded version of the dimensionless coupling constant $g$. Note that the retarded couplings with small momentum transfer $\tf{s}{ij}, \tc{s}{ij}$ do not flow at one-loop order.

We will assume in the following that Eqs. \eqref{rflow} remain valid
until the energy scale of the electronic spin/charge gaps is reached.
Physically, this corresponds to the assumption that the Debye scale
$\omega_{\mysw{D}}$ is of the order of the electronic gaps or less. The
evolution of the interactions will then be governed by the coupled set
\eqref{nrflow} and \eqref{rflow} of
nonlinear differential equations as
higher-energy electronic excitations are recursively integrated out. Since 
the nonretarded flow equations \eqref{nrflow} are unaffected by the retarded 
couplings,
we may regard their solutions as
known. Thus, all we need to do is to substitute them into
Eqs. \eqref{rflow} and solve for the renormalized retarded interaction
parameters.

It is rather remarkable that the RG equations \eqref{rflow} for retarded couplings are completey decoupled by introducing the following linear combinations of the original couplings, $\tilde{h} ^\pm_{ij}= \tf{\ell}{ij}\pm \tc{\ell}{ij}$. After the change of coupling basis, the RG equations take the simple decoupled form,
\begin{equation}
\partial_l \tilde h^\pm_{ij}=\rho^\pm_{ij}\,\tilde h^\pm_{ij}-2(\tilde h^\pm_{ij})^2,
\label{decoup}
\end{equation}
where $\rho^\pm_{ij}=(2\f{s}{ij}-4\f{\ell}{ij})\pm(2\coup{s}{ij}-4\coup{\ell}{ij})$. Note that, since $\tilde c_{ii} \equiv 0$, the label $\pm$ in \Eq{decoup} is redundant for $i=j$ and thus will be dropped.
Omitting indices, the solution of \Eq{decoup} is of the form
\be
\tilde h(l)= M(l)\left(\int_0^l dl'\, 2M(l')+\frac{1}{\tilde h(0)}\right)^{-1},
\label{sol}
\ee
where $M(l)=\exp\left(\int_0^ldl'\,\rho(l')\right)$. In the vicinity of the divergent scale $l_d$ where the nonretarded interactions diverge, the coefficient function $\rho(l)$ diverges as
\begin{equation}\label{asym}
  \rho(l)\sim \frac{\beta}{l_d-l},
\end{equation}
where the coefficient $\beta$ depends on the subscript $ij$ and
the superscript $\pm$ of $\rho$. Because of the above divergence
the retarded couplings will be {\it driven} to large values. Given
\Eq{asym}, the following three cases need to be distinguished:

(1) $\beta\geq 1$. The integral over $M$ in \Eq{sol} diverges.
The retarded couplings will then diverge as $(\tilde l_d-l)^{-1}$ with $\tilde l_{d} \leq l_d$. In the interesting case where the retarded interaction is attractive $\tilde h(0)<0$ (which is realized for $\tilde h^+_{ij}$ and $\tilde h_{ii}$), the quantity in the parenthesis in \Eq{sol} vanishes {\em before} the scale $l_d$ is reached, and hence $\tilde l_d<l_d$. In this case, retarded couplings eventually dominate over nonretarded ones, so long as $\omega_{\mysw{D}}$ is sufficiently small (compared with gaps in the system). For $\tilde h(0)>0$ we have $\tilde l_d=l_d$.

(2) $0< \beta <1$. A finite critical initial coupling
\begin{equation}\label{crit}
\tilde h_c=-\left(\int_0^{l_d}dl\,2M(l)\right)^{-1} <0
\end{equation}
can be identified. For $\tilde h(0)<\tilde h_c<0$, the behavior is
similar to case 1) with $\tilde l_d<l_d$. For $\tilde h(0)> \tilde
h_c$, $\tilde h(l)$ diverges as $(l_d-l)^{-\beta}$. This implies
that $\tilde h$ is still relevant, but is ultimately subdominant
compared to the nonretarded couplings, which diverge as
$(l_d-l)^{-1}$.

(3) $\beta<0$. Again, the behavior for $\tilde h(0)<\tilde h_c$ is
as in case 1) with $\tilde l_d<l_d$, but otherwise $\tilde h(l)$ is irrelevant and
vanishes as $(l-l_d)^{|\beta|}$.

At each point of the C1S0 phase of the pure Hubbard ladder,
the retarded couplings may now be classified according to the
above scheme. We shall now do so for the two-leg ladder. The
presentation of the results is greatly simplified by the fact that
the parameter $\beta$ in \Eq{asym} only depends on the ratio
of the two Fermi velocities $v_1/v_2$, where the indices $1$ and
$2$ refer to the majority (bonding) and minority (antibonding) bands, respectively.

\begin{figure}
\centering
\includegraphics[width=2.1in]{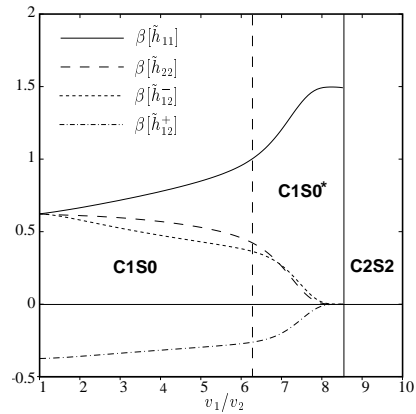}
\caption{\label{beta} $\beta$ parameters defined in \Eq{asym} for the retarded couplings of a two-leg ladder. 
Note that the intraband coupling $\tilde h_{11}$ divides the C1S0 phase into two regimes regarding whether phonons play the dominant role in determining the gaps. Numerically, $\beta$ has been obtained at $l_{c}$ for $U/t=10^{-5}$.}
\end{figure}

\begin{figure}
\centering
\includegraphics[width=2.1in]{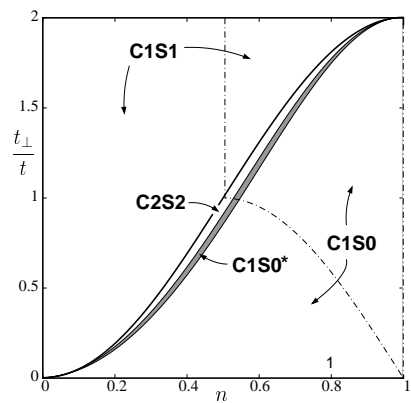}
\caption{\label{phas} Phase diagram of the two-leg Hubbard ladder as a function of $t_\perp/t$ and particle density. $t_\perp$ and $t$ are hopping amplitudes along rungs and chains.  Phases are denoted by CmSn, where m and n are the number of gapless charge and spin modes, respectively. In the regime below the thick line, both bands are partially occupied. The dash-dotted line indicates where umklapp processes are important (Refs. \onlinecite{BALFISHER,LINFISHER}) which were neglected in this work. 
 The velocity ratio $v_1/v_2$ equals 1 on the $t_\perp=0$ and $n=1$ axes,
and diverges towards the C1S1 phase boundary. }
\end{figure}

The $\beta$ parameters of the retarded couplings according to
\Eq{asym} are displayed in \Fig{beta}. As is readily seen, below $v_1/v_2\simeq 6$ in the C1S0 phase, the $\beta$ values for the most relevant retarded couplings behave according to the second case discussed above. The
critical initial values, $\tilde h_c$, for these couplings defined
in \Eq{crit} are of the order of the bare electronic on-site
repulsion $U$.\cite{Seidel} In  materials where the bare retarded
couplings are small compared to $U$, the majority of the fully
spin-gapped C1S0 phase (\Fig{phas}) will then have retarded
couplings diverge subdominantly, implying that
they {\em will not} surpass their electronic counterparts.
This finding is dramatically different from the conventional wisdom drawn from strictly one-dimensional systems,\cite{ZIMANYI,BINDLOSS} where even very weak retarded couplings will dominate the nonretarded ones and dictate the phase. On the other hand, the effects of phonons are, though less dominant, by no means negligible in the present scenario. This is particularly true for physical effects, such as the isotope effect, that primarily depend on the strength of the electron-phonon interaction: The renormalization of electronic properties due to phonons will be considerably stronger under circumstances where the
electronic interactions are relevant and open up gaps in the spectrum. 
This is best demonstrated by comparing the behavior described above for the C1S0 phase to that found in the C2S2 phase, where electronic interactions are
irrelevant. Note that, in the C2S2 phase (no spin gap), the quantity $\rho(l)$ will not show the behavior displayed in \Eq{asym}, but rather approach a constant value as $l\rightarrow\infty$. In this case, even though the retarded couplings may formally diverge for
$\rho(\infty)>0$, the energy scale associated with this divergence
{\em vanishes} as a power of the bare electron-phonon couplings.\cite{Seidel} They are thus irrelevant when the cutoff $\omega_{\mysw{D}}$ is
taken into account, below which Eqs. \eqref{rflow} are not valid.
Again we note that this dichotomy between
gapped and gapless phases is paralleled 
by the experiment of
Ref. \onlinecite{gweon}, where the reported isotope effect is much more pronounced
for temperatures and in regions of k space where a $d$-wave superconducting 
gap prevails.

The presence of phonons also has an influence on the spin gaps. First, we point out that in simple models the bare value of $\tilde h^-_{12}$ is usually zero owing to time-reversal symmetry. From \Eq{decoup}, $\tilde h^-_{12}$ will then remain zero within one-loop RG. $\tilde h^+_{12}$ is irrelevant for small initial values due to the negative sign of the corresponding $\beta$ parameter (\Fig{beta}). Under these conditions the relevant retarded couplings are {\em intraband} couplings. The latter become large and negative under RG, whereas their
electronic, initially repulsive, counterparts change sign under RG
and eventually also flow to large negative values. Hence attractive
retarded interactions tend to {\em enhance} the effects of
electronic interaction in causing the spin gap. In addition, the
increase of attractive interactions at large momentum transfer due
to electron-phonon coupling will tend to enhance charge-density wave (CDW) correlations, similar to the case of the single-chain system. 

In a slim regime of the phase diagram (above $v_1/v_2 \simeq 6$), the $\beta$ value of $\tilde h_{11}$ is larger than 1, such that the first case
mentioned above applies. In this narrow regime, the phonons take over the lead as in the single chain. This change can be roughly understood as the follows: Due to the large mismatch in Fermi velocities, the two bands decouple as two independent single-chain systems. In fact, in the regime above $v_{1}/v_{2} \sim 8$ (often misidentified as C2S1 phase in the literature), $\beta_{11} \approx 3/2$, which is the same as in the spin-gapped phase of a single chain. It is a bit tricky to determine the phase in this regime. Naively, at the cutoff scale $l = l_{c}$, the dominant retarded coupling $\tilde h_{11}(l_{c})$ reaches order $1$ and opens up the spin gap in the bonding band, while all other retarded and nonretarded couplings remain small. It is tempting to reach the wrong conclusion that the phase C2S1 is revived. Upon more careful analysis,\cite{EMERY, Seidel} the subdominant couplings $\tilde h_{22}, f^{\ell}_{22}$ also develop a minor spin gap in the antibonding band. In addition, the subdominant coupling $c^{\ell}_{12}>0$ pins the relative phases between two bands and leads to the ``$d$-wave'' pairing symmetry. Therefore, the final phase is again the same C1S0, except the spin gap in the bonding band is much larger than the others.


To conclude, we have developed a general RG scheme to analyze the effects of phonons in the correlated $N$-leg ladder. In particular, we demonstrate that the phonon-mediated attractive interaction, although divergent, remains subdominant compared to the electron-electron interactions in the two-leg ladder. This result
is particularly interesting when compared to that for the Luther-Emery
phase  of a single chain, where phonon effects can be even more profound. 
The general tendency suggested by these findings is that
as one moves away from a purely one-dimensional chain, 
the essential part in determining the phase diagram is played by
instantaneous interelectronic interactions. Phonons may, however, inject an
important bias when two leading competing electronic divergences are
equal in strength.
At the same time, the results obtained here highlight that electron-phonon 
couplings may have an enhanced influence
on quantitative aspects, in particular for measurements that probe energies
below the spin gap set by electronic interactions.


We thank A. Lanzara, Z.-X. Shen, and G.H. Gweon for insightful discussions. AS and DHL acknowledge the support of DOE grant DE-AC03-76SF00098. HHL is grateful for the financial support from Ta-You Wu Fellow and National Science Council in Taiwan through grants, No. NSC-91-2120-M-007-001 and NSC-93-2112-M-007-005.


\vspace{-5mm}

\begin{thebibliography}{10}

\bibitem{lanzara}
A. Lanzara {\it et al.}, Nature {\bf 412}, 510 (2001).

\bibitem{shen}
Z.-X. Shen, A. Lanzara, S. Ishihara and N. Nagaosa, Phil. Mag. B {\bf 82}, 1349 (2002).

\bibitem{gweon}
G.~H. Gweon {\it et al.}, Nature {\bf 430}, 187 (2004).

\bibitem{Rice03}
A. L\"auchli, C. Honerkamp and T.M. Rice,
Phys. Rev. Lett. {\bf 92}, 037006 (2004).

\bibitem{KIVELSON}

S.~A. Kivelson, Synthetic Metals {\bf 125}, 99 (2002)

\bibitem{LINFISHER}
H.-H. Lin, L. Balents and M.~P.~A. Fisher, Phys. Rev. B {\bf 56}, 6569 (1997).

\bibitem{FABRIZIO1}
M. Fabrizio, Phys. Rev. B {\bf 48}, 15838 (1993).

\bibitem{BALFISHER}
L. Balents and M.~P.~A. Fisher, Phys. Rev. B {\bf 53}, 12133 (1996).

\bibitem{SCHULZ}
H.~J. Schulz, Phys. Rev. B {\bf 53}, R2959 (1996).

\bibitem{ABRAHAMS}
G.~S. Grest, E. Abrahams, S.-T. Chui, P.~A. Lee and A. Zawadowski, Phys. Rev. B {\bf 14}, 1225 (1976).

\bibitem{CARON}
L.~G. Caron and C. Bourbonnais, Phys. Rev. B {\bf 29}, 4230 (1984).

\bibitem{ZIMANYI}
G.~T. Zimanyi, S.~A. Kivelson and A. Luther, Phys. Rev. Lett. {\bf 60}, 2089 (1988).

\bibitem{BINDLOSS}
I.~P. Bindloss, cond-mat/0404154 (unpublished).

\bibitem{SHANKAR}
R. Shankar, Rev. Mod. Phys. {\bf 66}, 129 (1994).


\bibitem{EMERY}
V.~J. Emery, S.~A. Kivelson and O. Zachar  Rev. B {\bf 59}, 15641 (1999).

\bibitem{Seidel}
A. Seidel, H.-H. Lin and D.-H. Lee, work in progress.

\end{thebibliography}
\end{document}